\documentclass{aastex}
\usepackage{spr-astr-addons}    

\usepackage{multirow}
\usepackage{multicol}
\usepackage{graphicx}	
\usepackage{amsmath}	
\usepackage{amssymb}	
\usepackage[dvipsnames]{xcolor}
\usepackage{array,tabularx}
\usepackage[colorlinks,allcolors=blue]{hyperref}
\usepackage{natbib}

\usepackage{soul}
\usepackage{hyperref}
\newcommand{\be}{\begin{equation}}
\newcommand{\ee}{\end{equation}}
\newcommand{\bea}{\begin{eqnarray}}
\newcommand{\eea}{\end{eqnarray}}

\begin{document}

\title{Web of resonances and possible path of evolution of the small Uranian satellites}

\shorttitle{Resonances and evolution of small Uranian moons}
\shortauthors{C. Charalambous et al.}

\author{C. Charalambous \altaffilmark{1,2}} 
\and 
\author{C.A. Giuppone \altaffilmark{2}}
\and 
\author{O.M. Guilera \altaffilmark{3,4}}
\email{carolina.charalambous@unamur.be} 

\altaffiltext{1}{naXys, Department of Mathematics, University of Namur, Rue de Bruxelles 61, 5000 Namur, Belgium}
\altaffiltext{2}{Universidad Nacional de Córdoba, Observatorio Astron\'omico - IATE. Laprida 854, 5000 C\'ordoba, Argentina}
\altaffiltext{3}{Instituto de Astrof\'{\i}sica de La Plata, CONICET-UNLP, Paseo del Bosque S/N, 1900 La Plata, Argentina}
\altaffiltext{4}{N\'ucleo Milenio de Formaci\'on Planetaria (NPF), Chile}
	
\begin{abstract} 
Satellite systems around giant planets are immersed in a region of complex resonant configurations. Understanding the role of satellite resonances contributes to comprehending the dynamical processes in planetary formation and posterior evolution. Our main goal is to analyse the resonant structure of small moons around Uranus and propose different scenarios able to describe the current configuration of these satellites. We focus our study on the external members of the regular satellites interior to Miranda, namely Rosalind, Cupid, Belinda, Perdita, Puck, and Mab, respectively. 
We use N-body integrations to perform dynamical maps to analyse their dynamics and proximity to two-body and three-body mean-motion resonances (MMR). We found a complicated web of low-order resonances amongst them. 
Employing analytical prescriptions, we analysed the evolution by gas drag and type-I migration in a circumplanetary disc (CPD) to explain different possible histories for these moons. We also model the tidal evolution of these satellites using some crude approximations and found possible paths that could lead to MMRs crossing between pairs of moons. Finally, our simulations show that each mechanism can generate significant satellite radial drift leading to possible resonant capture, depending on the distances and sizes.
\end{abstract}



\section{Introduction}
The Solar System is an abundant lab in which theories of planet and satellite formation and evolution can be developed and tested. The similarity of the giant planets of our Solar System and their moons with Kepler-like systems permits an analogy, and it is possible to analyse whether the same theories are valid in both cases.
Densely packed systems of regular satellites allow us to constrain the formation theories and predict how they achieved their observed configurations.

Planets are supposed to migrate to their current positions during the planetary formation process after interacting with the protoplanetary disc \citep{GoldreichTremaine1979,LinPapaloizou1979,2002ApJ...565.1257T,Paardekooper+2011,JimenezMasset2017}, followed by tidal dissipation from the star \citep{2001ApJ...548..466B,2013AJ....145....1B}. The expected result is that adjacent pairs of planets could be captured in mean-motion resonance (MMR). Thus, orbital commensurabilities are an essential mechanism in shaping (the dynamics of) planetary systems, and, if the analogy is correct, it should also be the case for satellite systems. 
However, \citet{Petrovich+2013} developed a model of in-situ formation with mass accretion that could also explain the period-ratio distribution among exoplanets. 

There are more than enough examples of satellite commensurabilities within the Solar System. Jupiter's Galilean satellites Io, Europa, and Ganymede, are captured in 2 and 3 body resonances. Commensurability relations are 2/1 between Io and Europa, 2/1 between Europa and Ganymede, and a Laplace type relation involving the three bodies \citep{1981Icar...47....1Y,1998A&AS..129..205L}. Something similar is observed in Pluto, where Hydra, Nix, and Styx form a 3:5:2 resonant chain around the binary Pluto-Charon system \citep{2015Natur.522...45S,2018DPS....5022108D}. Two-body resonances are also observed in the Saturn satellite system \citep{Sinclair1972,MeyerWisdom2008}. The pairs Enceladus-Dione and Mimas-Tethys are in a 2/1 MMR, while Titan-Hyperion present a 4/3 commensurability.
Naiad and Thalassa, Neptune's inner moons, appear to be locked in the 73/69 fourth-order orbital resonance, while Hippocamp and Proteus are in the 13/11 second-order MMR \citep{2020Icar..33813462B}. Uranus satellites are no exception, Belinda and Perdita appear to be librating in the 44/43 MMR\footnote{For a detailed discussion see Section~\ref{appendix}.} \citep{French+2015}, and Miranda-Ariel are very close to a 5/3 MMR \citep{Tittemore+1989}, but we will discuss this system in more detail in the following. 
The observed resonant configurations may be explained through orbital evolution in a primordial disc surrounding the planets, consistent with migration followed by resonant capture and posterior tidal evolution \citep{MosqueiraEstrada2003,Crida+2012}. 

William Herschel first discovered Uranus in 1781. Since then, 27 natural moons have been detected orbiting the planet: regular and external irregular ones. 
The regular satellites are composed of two different groups, the lesser moons and Miranda, Ariel, Umbriel, Titania, and Oberon. These last five are usually known as the major classical moons.\footnote{Ariel and Umbriel were both discovered by Lassell in 1851, Titania and Oberon by Herschel in 1787 and lastly Miranda in 1948 by Kuiper. See \href{https://planetarynames.wr.usgs.gov/Page/Planets}{https://planetarynames.iau}.} 
The internal group forms the most densely packed system of low-mass satellites in the Solar system. Their members are Cordelia, Ophelia, Bianca, Cressida, Desdemona, Juliet, Portia, Rosalind, Cupid, Belinda, Perdita, Puck, and Mab.\footnote{The Voyager 2 team \citep{Smith+1986} discovered 10 of the 13 moons of the inner regular satellites. Perdita was first reported by \citet{Karkoschka2001}, while Mab and Cupid were detected by \citet{Showalter+2006} }  
Both regular groups are located deep inside the irregular set.

Uranus spins on its side, and its satellite system orbit equally inclined. The two most common explanations are (i) that this system is formed as a consequence of an impulsive giant impact \citep{1966SvA.....9..987S}, or (ii) the complete system slowly shifted as a whole due to a resonance between the precession rates of the spin axis and of the orbit \citep{2010ApJ...712L..44B}.
Uranian regular moons are supposed to have formed either from a post-impact debris disc \citep{1992Icar...99..167S,Kegerreis+2018} or from a pre-impact proto-satellite disc that was destabilized by the post-impact debris disc and rotated to become equatorial \citep{2006Natur.441..834C,Morbidelli+2012}. \citet{ida2020}, on the other hand, propose that the Uranian satellite formation is regulated by the evolution of the impact-generated disc. 

Many studies were devoted to understanding the dynamics of the classical satellites. \citet{1984A&A...140...33L} studied the Laplace resonances between Miranda-Ariel-Umbriel. 
\citet{Tittemore+1988} work provided analytical treatment of Uranus' classical satellites, which \citet{Cuk+2020} extended for studying their past tidal evolution. According to their work, the primary interaction in the system was between Ariel and Umbriel when they crossed the 5/3 mean-motion resonance, and the currently observed eccentricities and inclinations within the whole system are due to secular resonances.

\begin{table*}
\centering
\caption{Semi-major axes, eccentricities, masses, and radii of the inner regular satellites. $a$ and $e$ values taken from the JPL Solar System Dynamics database (\href{https://ssd.jpl.nasa.gov/horizons}{https://ssd.jpl.nasa.gov/horizons}). Horizontal line separates the inner and outer members of the group. All radii taken from \citet{Showalter+2006} except from Cordelia and Ophelia taken from \citet{Karkoschka2001}. Masses taken from \citet{French+2015}.  \label{tab:a}}
\begin{tabular}{l|c|c|c|c}
    & $a_{\rm URA115}$ & $e_{\rm URA115}$ &  $mass$ & $radius$ \\
    \multirow{-2}{*}{Satellite} & $[{\rm km}]$ & $[10^{-3}]$ & $[10^{16} \, {\rm kg}]$ & $[{\rm km}]$\\
    \tableline
        Cordelia & 50028.789 & 4.27 & 3.87924 & 20.1 \\
        Ophelia & 54154.072 & 16.22  & 5.09650 & 21.4 \\
        Bianca & 59344.546 & 3.18  & 8.24480 & 27    \\
        Cressida  & 61844.078 & 0.80 & 28.8696 & 41  \\
        Desdemona & 62686.574 & 0.15  & 17.9594 & 35  \\
        Juliet & 64442.560 & 1.83 & 62.3615 & 53 \\
        Portia & 66141.888 & 0.48  & 143.676 & 70  \\
    \tableline
        Rosalind & 70018.176 & 1.22 & 19.5432 & 36  \\
        Cupid & 74574.313 & 6.27  & 0.295297 & 8.9 \\ 
        Belinda & 75324.586 & 0.66 & 31.8704 & 45  \\ 
        Perdita & 76586.518 & 3.51  & 0.985470 & 13.3 \\ 
        Puck & 86077.334 & 0.59 & 222.609 & 81  \\ 
        Mab  & 97752.516 & 3.31 & 0.79865  & 12.4 \\
    \tableline
\end{tabular}
\end{table*}
The inner regular satellite system of Uranus had been extensively studied by numerical means, and strong gravitational instability is predicted among them. Multiple observations show a significant variation of the semi-major axes of the inner satellites in timescales of decades. For example, \citet{Duncan+1997} found that the five major satellites were stable for longer than the age of the solar system, while the inner satellites were stable over a much shorter period ($\sim 4-100$ million years). \citet{Showalter+2006} argued that the variations in the orbital elements might be a short-term manifestation of the predicted long-term instability. They showed that the instability is due to multiple mean-motion resonances between pairs of satellites and predicted that Cupid-Belinda or Cressida-Desdemona have crossing orbits. \citet{French+2012} also investigated the sensitivity to small changes in initial conditions and explored the role of resonances in causing the long-term instability of the system.

\citet{2011MNRAS.418.1043Q} and \citet{Quillen+2014} explored the dynamics of resonant chains within the Portia satellites through the 3-satellite eccentricity-type resonances, analogous to the Laplace resonance involving Io, Ganymede, and Europa.
\citet{Quillen+2014} reported that the strongest three-satellite commensurability between Cressida, Desdemona, and Portia is 46:-57:13, near the 46/47 first-order MMR between Cressida and Desdemona and the 12/13 MMR between Desdemona and Portia. Such high-order MMR are weak, even considering that the mass ratio between Uranus and its moons is lower than $10^{-5}$ and in nearly coplanar and circular orbits.

Motivated by the observed orbital changes in the inner Uranian moons, \citet{French+2015} carefully investigated the interlinked resonances among the Portia group, inclined and in the orbital plane. They explored their mutual gravitational interactions to reveal the short-term manifestations of the destabilising resonant interactions that can eventually lead to crossing orbits and understand the conditions leading to orbital chaos. They consider that the moons' dynamical couplings cause both the regular and irregular variations in their orbital elements, depending on the assumed satellite masses.

Here, we follow and extend the previously mentioned ideas to explain the dynamics behind the lesser satellites' distribution orbiting around Uranus. Our focus is on the small external satellites to Portia semi-major axis, i.e, the last six below the horizontal line in Table~\ref{tab:a}, mainly because they were not analysed before, and because Belinda and Puck are the most massive moons of the internal group, so it is to be expected that their dynamical importance is not negligible. We aim to qualitatively understand the resonant structure in the different planes, and propose a realistic scenario of how the satellites arrived at their current location. It is important to stress that we do not intend to propose an origin for the CPD nor the moons. Our concern is to understand their stability and how they achieved their current configuration. Thus, we follow the work of \citet{ida2020} as the framework for the possible formation of the Uranian minor moons, which could have formed in the CPD and suffer the typical interactions with it, mainly the gas drag due to their sizes and small masses.
Table \ref{tab:a} shows relevant semi-major axes, eccentricities, masses, and radii of the moons. We obtain the ecliptical orbital elements from the Jet Propulsion Laboratory for the Solar System Dynamics (JPL) Horizons database, at epoch 2021/01/01 (\href{https://ssd.jpl.nasa.gov/horizons}{https://ssd.jpl.nasa.gov/horizons}). We used planetocentric orbital elements corresponding to the pre-computed URA115 solution.

We organise this work as follows. First, we describe the dynamical maps in Section \ref{sect:dyn-maps}, which help us understand the resonant structure around Uranus. We focus on the small outer satellites of the regular satellites interior to Miranda's orbit and describe their complexity in Section \ref{sec:dyn_maps}. Sections \ref{sect:disc-capture} and \ref{sect:tidal} provide a possible evolution for the satellites, comparing capture in low order MMR due to disc-driven migration with tidal interactions with the central planet. Finally, we present a brief discussion and our conclusions in Section \ref{sec:conclusions}.

\section{Resonant structure considering three moons} \label{sect:dyn-maps}
\begin{figure}
\centering
\includegraphics[width=0.9\columnwidth]{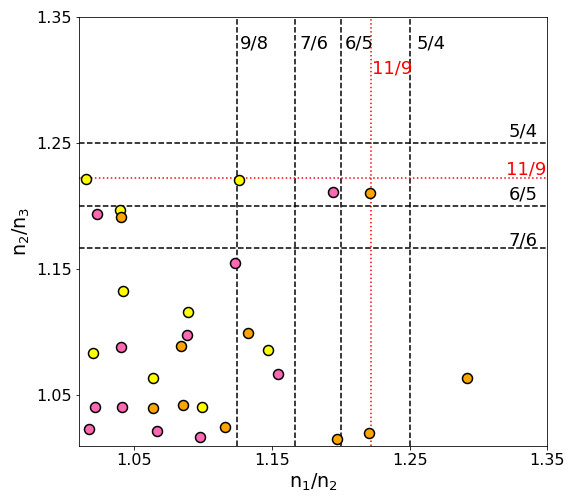}
\caption{Satellite distribution in the $(n_1/n_2,n_2/n_3)$ plane. Dots with different colours show different sets of the regular internal satellites taking three moons in ascending semi-major axis, see Table~\ref{tab:a}. Pink represent three consecutive moons (e.g., Cordelia, Ophelia, Bianca). Orange represent 3 moons in order skipping the second moon in the order (e.g. Cordelia, Bianca, Cressida) and yellow dots considering the sets skipping the third moon in the order (e.g., Cordelia, Ophelia, Cressida). Vertical and horizontal lines depict the main resonances.}
\label{fig:mmr}
\end{figure}
\begin{figure}
\centering
\includegraphics[width=\columnwidth]{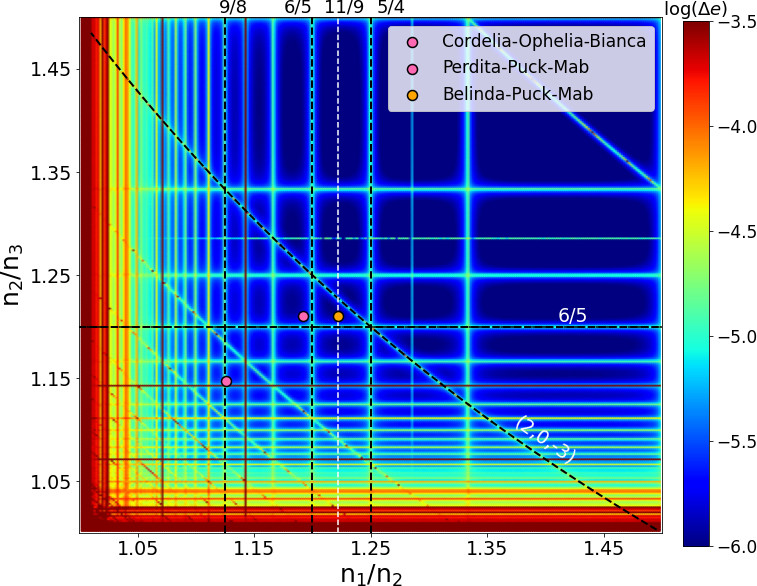}
\caption{$max (\Delta e)$ dynamical map for three equal mass satellites around Uranus for $300 \times 300$ initial conditions. All masses were taken as the bigger moon masses ($m_i \simeq 10^{18} \, {\rm kg}$). Resonances $p+q/p$ appear as vertical and horizontal lines. Relevant commensurabilities in the Uranian system are highlighted as dashed white curves. Between inner and middle bodies, from left to right, 9/8, 6/5 and 5/4. We also stressed the 11/9 MMR, although it is not seen in the background map. 6/5 MMR between middle and external moons is showed as an horizontal line, and the diagonal curve represent the 2/3 commensurability between the non-adjacent satellites. Superimposed to the representative map and following the same colour code as in Figure~\ref{fig:mmr}, the triplets near the main commensurabilities.}
\label{deUr}
\end{figure}
In this section, we consider Uranus as the massive central body and three of its satellites. In our numerical simulations, we integrate the equations of motion of the four bodies of the system in a planetocentric reference frame. We perform the computation using a Bulirsch–St\"oer algorithm, the most suitable for reproducing close encounters while preserving the topology of the orbits. We set the mass and radius of Uranus $M_U = 4.365 \times 10^{-5} \, {\rm m}_\odot$, and ${\rm R_U} = 25600 \, {\rm km}$, and we chose to fix the inner mean-motion $n_1$ equal to that of Belinda. We fixed the eccentricities in $e_i = 0.001$, and use $\lambda_1= \lambda_2 = \lambda_3 = 0$. At this stage, we did not include any effects of the disc, such as migration or tidal interaction. 
The simulation finishes when the integration reaches 500 years or when a collision or escape takes place. We qualified a collision with the central planet as the minimum distance lower than $2\times 10^{-4} \, {\rm au}$ (i.e., $\sim 1.1 \, {\rm R_U}$). An escape is considered when the distance from the primary is bigger than 0.2 $\, {\rm au}$. A collision occurs if the distance between two bodies is smaller than the sum of their physical radii $R_i+R_j=1$. Eccentricities bigger than 0.99 also lead to escapes from the system. 

We explore the dynamics of the three small moons in the $(n_1/n_2,n_2/n_3)$ representative plane \citep[see .e.g., ][]{Migaszewski+2016,Charalambous+2018,Petit2021}. In the colour scale, we use either ${\rm max}(\Delta a)$ or ${\rm max}(\Delta e)$ indicators. Although ${\rm max}(\Delta a)$ and ${\rm max}(\Delta e)$ are not chaos indicators, they are essential tools widely used among the dynamical community to identify the resonant structures, i.e., the positions of stationary solutions as well as the separatrix of different commensurabilities \citep[see, for example,][]{2004A&A...426L..37D,2015CeMDA.123..453R,2017A&A...602A.101R,Charalambous+2018}. 

A dynamical system consisting of two masses orbiting a central body is in a mean-motion resonance $(p+q)/p$ when the mean-motions $n_i$ of the tiny bodies under consideration satisfy the relation $(p+q)n_2 - p n_1\sim 0 $ with $p,q \in \mathbb{Z}$. The dominant term in the inner satellite's disturbing function has a resonant argument  $\phi_{1,2} \sim (p+q)\lambda_2 -p \lambda_1 -q\varpi_{1,2}$, with $\lambda_i$ the mean longitudes and thus $\varpi_i$ the longitudes of pericentres. Following \citet{Morbidelli2002} and \citet{2015CeMDA.123..453R}, we will refer to $p$ as the degree and $q$ as resonance order, respectively.

\begin{figure*}
\centering
\includegraphics[width=0.9\textwidth]{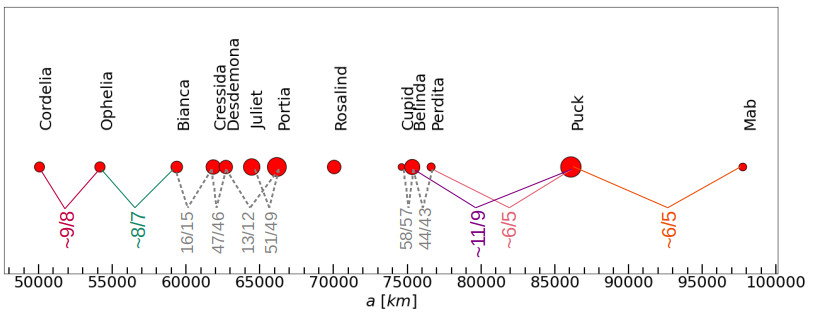}
\caption{Schematic representation of the closeness of consecutive period ratios to ratio of integers among the small moons of Uranus, illustration inspired in \citet{French+2015}. Dashed grey lines show previously reported resonances by \citet{French+2015} for the Portia group, while in continuous colour lines we show the proximity to lower order resonances recognised in Figure~\ref{deUr}. See text for detailed discussion.}
\label{res-french}
\end{figure*}

In Figure~\ref{fig:mmr}, we show the lesser satellite distribution in the $n_1/n_2,n_2/n_3$ representative plane. We use the JPL Solar System Dynamics database to compare the dynamic structures and identify resonances in the different systems. Each dot represents a triad of moons of the regular small satellite group analysed in different ways. In the extent of the $(n_1/n_2,n_2/n_3)$ plane plotted, only lesser satellites are observed. The classical moons are out of range. Pink dots represent three consecutive moons. Yellow and orange show groups of four, having skipped the second or the third body of the set, respectively. Although this way of considering the different interactions does not seem straightforward, it allows us to recognise resonances proximity between adjacent and non-adjacent pairs.  We also highlight the most important resonances visible in the plane. Vertical and horizontal dashed lines represent first-order MMRs, while the red dotted vertical line represents a second-order resonance, the 11/9.

An integration for 500 years over a grid of $300 \times 300$ initial conditions for three equal mass satellites is shown in Figure~\ref{deUr}. We consider $m_i=1\times 10^{18} \, {\rm kg}$, which roughly corresponds to Belinda's mass\footnote{Maps with different masses ($m_i = 10^{-11}, 10^{-13},$ and $10^{-15}\, {\rm m}_\odot$) show the same structure with both the ${\rm max}(\Delta a)$ and ${\rm max}(\Delta e)$ indicators.}. The position of the innermost satellite $a_1\sim 75000 \, {\rm km}$ is maintained fixed and we vary $a_2$ and $a_3$, consequently modifying the mean-motion ratios $n_1/n_2$ and $n_2/n_3$. The colour code corresponds to the maximum values attained by the eccentricities of the satellites, ${\rm max}(|e_i(t)-e_i(t=0)|)$. Red parts of the map indicate high variations in the eccentricity, while blue is associated with minor variations. Similar structures are also observed in the ${\rm max}(\Delta a)$ dynamical maps. 
The main features of the dynamical structure observed in the maps are the same as those presented in \citet{Charalambous+2018}: (i) vertical lines identify resonances involving two bodies $m_1$ and $m_2$; or, by symmetry, horizontal lines represent MMRs between $m_2$ and $m_3$. (ii) MMRs between the non-adjacent planets $m_1$ and $m_3$, are observed as diagonal curves. The intersection points between two 2-planet commensurabilities are {\it double resonances}. The structures revealed by this map help us identify the strongest or most important interactions in different configurations of three moons around Uranus. Superimposed to the dynamical map, we plotted the triplets within the regular internal satellite group, taken as different sub-sets, following the colour code of Figure~\ref{fig:mmr}.

\begin{figure}[h]
\centering
\includegraphics[width=0.95\columnwidth]{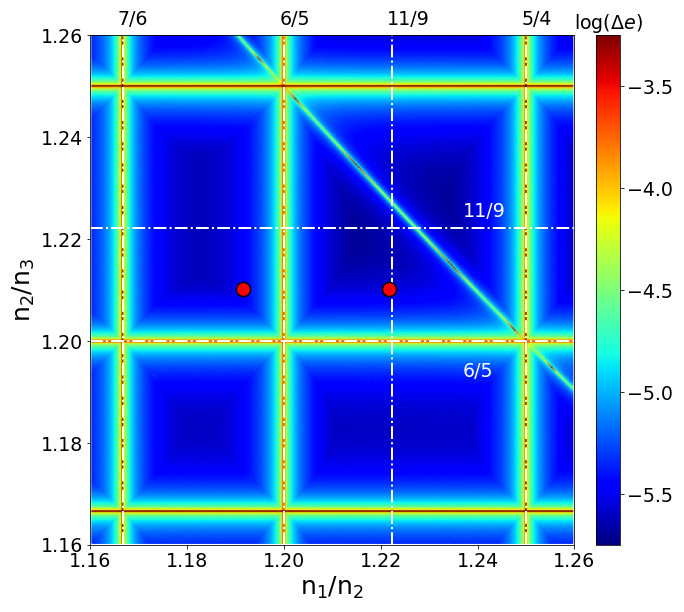}
\caption{Zoom in the ${\rm max} (\Delta e)$ of Figure \ref{deUr} in the $(n_1/n_2,n_2/n_3)=(1.16,1.26)$ region, constructed with the estimated masses of Perdita, Puck and Mab. Red dot at $(n_1/n_2,n_2/n_3)=(1.1915,1.210)$ stands for Perdita/Puck and Puck/Mab, while red dot at $(n_1/n_2,n_2/n_3)=$ $(1.221,1.210)$ stands for Belinda/Puck and Puck/Mab. }
\label{deUr_zoom}
\end{figure}

In Figure \ref{res-french}, we present the nearest dominant 2-satellite MMRs exhibited in Figure~\ref{deUr}, comparing with those identified in \citet{French+2015} for the Portia group, shown in continuous colour lines, and in grey dashed lines, respectively. Red dots show the position of the moons with the semi-major axes published in JPL. Although we also identify resonances of order $q=1$ and $q=2$, in \citet{French+2015} $p$ is much higher. For example, the 47/46 is a first-order ($q=1$) $46-th$ degree resonance. When we consider all such high-degree commensurabilities, they overlay between them, creating an unstable domain \citep[see Figure 6 from][]{2015CeMDA.123..453R}.

To better understand the resonant web in the satellites, Figure \ref{deUr_zoom} show a zoom for $n_1/n_2 \in (1.16,1.26)$, and $n_2/n_3 \in (1.16,1.26)$, considering the masses $m_i$ those of Belinda, Puck, and Mab, respectively. The innermost moon was set at $n_1=n_{\rm Belinda}$. Again $e_i=0.001$ and $\lambda_i=0$. Two different triplets are shown. The outer pair is always Puck-Mab ($n_2/n_3\sim 1.21$), near the $6/5$ resonance ($n_2/n_3 = 1.20$). The inner pair at the left represent Perdita-Puck, close to the $6/5$ MMR. The pair Belinda-Puck ($n_1/n_2\sim 1.221$) appears near the $11/9$ MMR ($n_2/n_3\sim 1.222$), although this resonance does not appear in the map. As the dynamical map is constructed for nearly circular orbits, a second-order commensurability such as the 11/9 is not expected to appear in this plane. 

In the next section, we construct dynamical maps and analyse the resonant structure in the $(a,e)$ plane for the outer moons within the regular low-mass satellites, internal to the Classical group, using the maps in Figures~\ref{deUr} and \ref{deUr_zoom} as guides.

\section{Dynamical maps}
\label{sec:dyn_maps}

We analyse the structure of dynamics between two satellites and three satellites in the following subsections to understand the proximity of the moons to the MMR identified in the previous section. The zonal harmonic $J_2$ term is a dominating perturbation term of Uranus's gravitational field, affecting those satellites closer to the planet more efficiently. Thus, we consider our set of simulations with and without this contribution to understand its effect in shaping resonances. 
The integrations do not include the five classical moons (Miranda, Ariel, Umbriel, Titania, and Oberon) as they do not influence the stability of the inner moons \citep{Duncan+1997, French+2012}. 

In this Section, we present numerical integrations in the $(a,e)$ plane to emphasise some dynamical features that could not be seen in the $(n_1/n_2,n_2/n_3)$ plane. In doing so, we use kilometres as the unit so that we can display the moons with the positions given by JPL (Table~\ref{tab:a}) and compare them with the resulting maps.

\subsection{Dynamical maps considering two moons}\label{sec:2-moons}

\begin{figure*}
\centering
\includegraphics[width=0.45\textwidth]{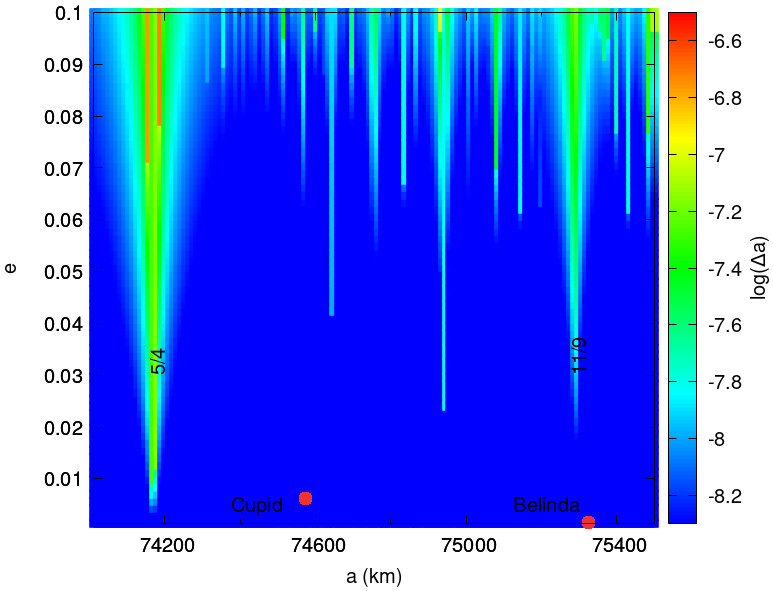} \quad
\includegraphics[width=0.45\textwidth]{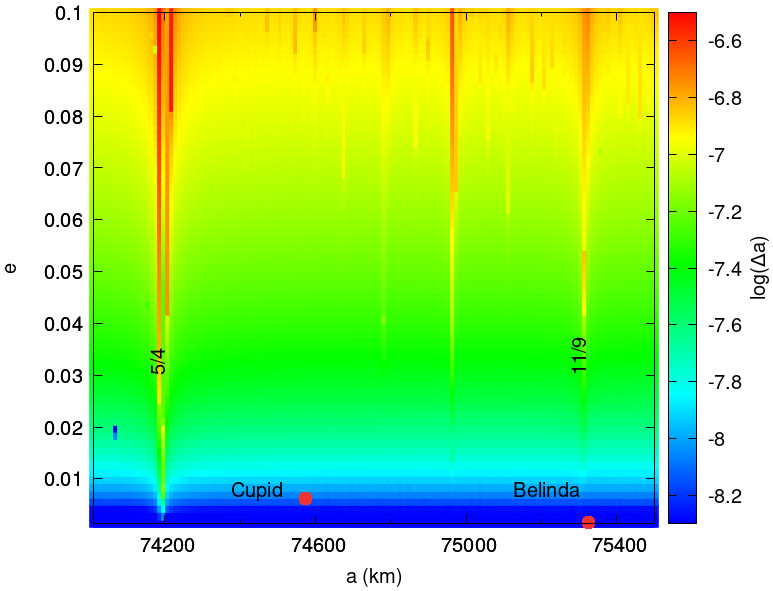}
\caption{\emph{Left panel:} Dynamical map of the internal regions to Puck in the $(a,e)$ plane using ${\rm max} (\Delta a)$ indicator, considering an additional small moon with Belinda's mass. The integration time is equal to $100 \, {\rm yrs}$. 5/4 (left) and 11/9 (right) MMRs are identified as vertical structures at $a\sim 74175 \, {\rm km}$ and $a\sim 75294 \, {\rm km}$, respectively. Red circles represent the positions of Cupid (left) and Belinda (right) according to the JPL. \emph{Right panel:} the same integration but considering the effect of Uranus' oblateness, $J_2$. }
\vspace{-1 cm}
\label{map-puck}
\end{figure*}

In Figure \ref{map-puck}, we present an integration for 100 years\footnote{The choice of the integration timescale was reduced from that of Figure~\ref{deUr} since at least one secular period of the eccentricity is covered, and we can shorten computational time.} over a grid of 100$\times$100 initial conditions, where we consider the moon Puck and an additional companion with the mass of Belinda.
We show the structure in the $(a,e)$ plane ranging the semi-major axis that covers the position of Cupid ($a\sim 74400 \, {\rm km}$) and Belinda ($a\sim 75300 \, {\rm km}$), varied the eccentricities $e_i \in [0,0.1]$, and fixed all initial angular variables to zero. 
From the dynamical maps in Figures~\ref{deUr} and \ref{deUr_zoom}, we can see the pair Belinda-Puck near the 11/9 MMR but, as already mentioned, this resonance is not observed in the $(n_1/n_2,n_2/n_3)$ plane. However, in Figure~\ref{map-puck} it is possible to observe the effect of the 11/9 MMR in the $(a,e)$ plane. The left frame shows a dynamical map without the $J_2$ zonal harmonic of Uranus, and in the right frame the zonal harmonic is considered. In left frame of Figure~\ref{map-puck} we observe
the V-shape of the 5/4 and 11/9 MMRs at $a\sim 74175 \, {\rm km}$ and $a\sim 75294 \, {\rm km}$, respectively. 
When we superimpose the current position of Cupid (left) and Belinda (right) (red dots) with their semi-major axes and eccentricities given in Table~\ref{tab:a} to compare their configuration with the background map, it is evident that Belinda is located close to the 11/9 commensurability. Cupid, however, lies clearly outside the libration domain of the 5/4 MMR, i.e., outside the observed V-shape that represents the separatrix of the resonance. When considering the effect of Uranus' oblateness, the dynamical maps exhibit the same structure and slightly shift the resonance location (see the right hand frame of Figure~\ref{map-puck}). In particular, the 11/9 MMR with Puck, which is located in the proximity of Belinda. The net effect of the zonal harmonic \citep[$J_2=3.343 \times 10^{-3}$,][]{1996P&SS...44...65A} 
implies more significant variations in the semi-major axis and eccentricities, and the structure in the resonance is more highlighted using the ${\rm max}(\Delta a)$ indicator. 

\subsection{Dynamical maps considering three moons}
\label{sec:3-moons}
\begin{figure*}
\centering
\includegraphics[width=0.45\textwidth]{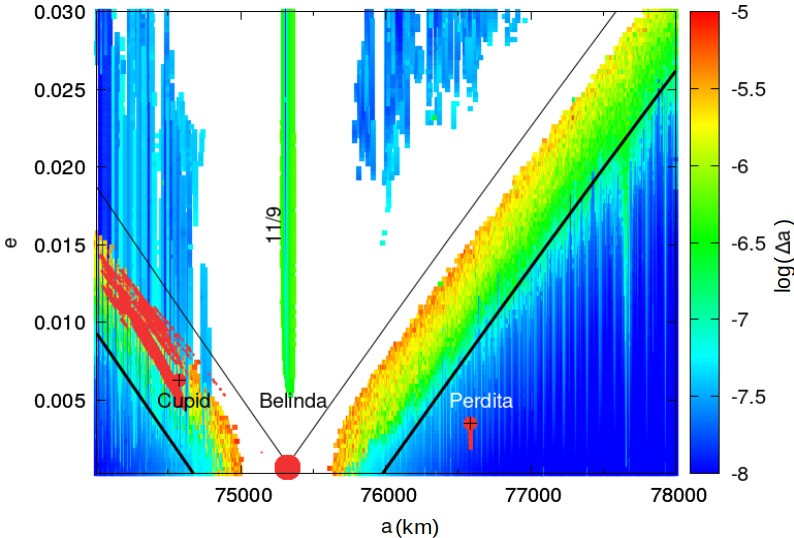} \quad
\includegraphics[width=0.45\textwidth]{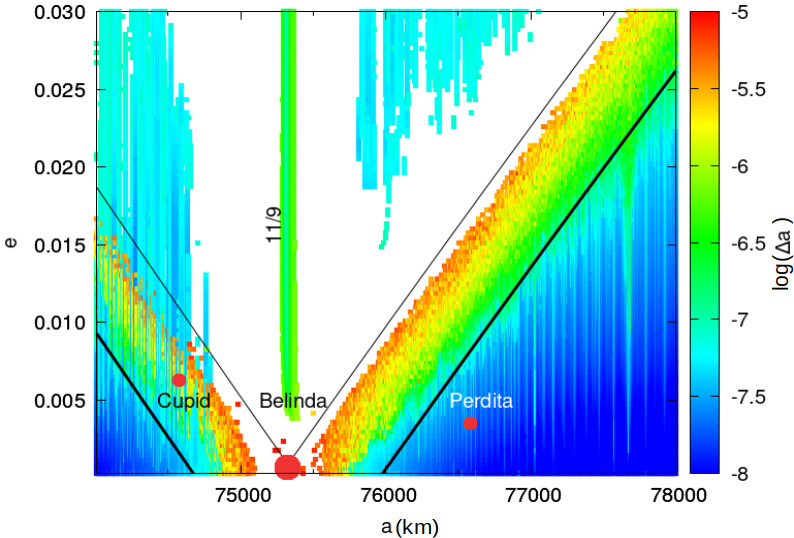}
\caption{\emph{Left panel:} $\, {\rm max} (\Delta a)$ dynamical map for three moons around Uranus for 100$\times$100 initial conditions in the regions around Belinda, not considering the oblateness effect (i.e., $J_2=0$). We mark with red circles the positions of Cupid and Perdita. The integration time is $200 \, {\rm yrs}$. Vertical line corresponds to the 11/9 MMR with Puck at $a \sim 75298\, {\rm km}$. Resonances appear as vertical structures providing stable motion above secular stability limits, identified as black inclined lines. White parts represents initial condition that did not finish the simulation because the moons collide or escape. Integrations of the N-body problem are overlapped, with black crosses indicating initial conditions. See the text for more detail. \emph{Right panel:} Same plot but considering the zonal harmonic $J_2=3343.43 \times 10^{-6}$. }
\label{Bel-Puck-int}
\vspace{-1 cm}
\end{figure*}

Further on, we consider the most massive moons, Belinda and Puck, and an additional moon with the mass of Cupid. Semi-major axes and eccentricities for the test particle are taken from a regular grid of 100$\times$100 initial conditions, and integrated with the Bulirsch–St\"oer algorithm for 200 years or, equivalently, more than 75 thousand revolutions of Mab around Uranus. Results are shown in Figure \ref{Bel-Puck-int}, where we analyse the resonant structure with the semi-major axis in the range $74000 \, {\rm km} <a< 78000 \, {\rm km}$. The moons were considered initially in coplanar orbits, varied the eccentricities between 0 and 0.03, and fixed all initial angles $\omega_i=0, \Omega_i =0$ and $M_i = 0$. As in the previous Section, the left hand side of Figure~\ref{Bel-Puck-int} considers integrations with $J_2=0$, while right hand side includes in the integration the current zonal harmonic of Uranus, $J_2 = 3.343 \times 10^{-3}$.
A crowded region of high-order resonances is present. We calculate the semi-empirical crossing orbit stability criterion for eccentric planetary systems, based on Wisdom's overlap criterion for first-order MMRs \citep{Giuppone+2013}, adapted to the Uranian system. The analytical expressions provide regions of stability. For eccentric orbits, the stability limits follow the pericentric and apocentric collision lines, shown as thin lines in the plot, starting from Belinda's position. We use the interior and exterior limit criterion for systems with $e>e_i$ (being $e_i$ the perturber's eccentricity, which is generally smaller than the test-moon eccentricity, $e$), and plot the lines superimposed to the dynamical map shown in Figure \ref{Bel-Puck-int} as solid thick black lines. 
Explicitly, when the two orbits are initially aligned (which is the case since all angles are set to zero at the beginning of the integration), the interior limit for orbit crossing for eccentricity of test moon, $e$, is greater that the perturber' moon, $e_i$, is given by
\begin{equation}
    a > (a_i - \delta_i) \frac{1+e_i}{1+e},
\end{equation}
and the exterior limit is given by
\begin{equation}
    a < (a_i + \delta_i) \frac{1-e_i}{1-e},
\end{equation}
where the quantity $a_i$ is the perturber's semi-major axis and $a$ is that of the test-moon. $\delta_i = [\mu_i^{2/7} + \mu^{2/7}]a_i$ represents a region of instability around the satellite, and $\mu$ is the mass ratio between each small moon with Uranus \citep[for details, see][]{Giuppone+2013}. 
The thick lines delimit the extended crossing orbits while the thinner ones represent the collision region. Orbits above the lines correspond to unstable regions. Note that both Cupid and Perdita lie close to the unstable apocentric and pericentric limits, respectively. We observe high values of ${\rm max} (\Delta a)$, green and red points above the curves and only survive vertical regions associated with high-order MMRs. The regions with $e>0.005$ at the 11/9 MMR with Puck ($a \sim a_{\rm Belinda}$) give some space for co-orbital companions to Belinda. Accordingly, this figure gives information about the richness of the resonant structure and the possible existence of higher-order resonances (see also Section \ref{appendix}).  

We study the stability of the initial conditions using four moons (Cupid, Belinda, Perdita, Puck) and over-plot the dynamical behaviour in the same map (see the red points in the left hand frame of  Figure~\ref{Bel-Puck-int}). We integrate the equations of motion setting the initial orbital elements given by JPL-Horizons at epoch 2021/01/01. Integrations show instability around $5\times 10^4 \, {\rm yrs}$ for JPL's initial conditions. Despite the fact that Perdita's eccentricity is of same order than Cupid's, Cupid evidence more irregularities and greater excursions in eccentricity more rapidly than Perdita. The smooth evolution of eccentricity of the small moons seems to reflect the diffusion observed in \citet{Gallardo+2012}. In other words, the integration show that Cupid is doomed in only 50000 years and not survive in the system.

The right frame of Figure~\ref{Bel-Puck-int} shows a regular grid in the $(a,e)$ plane, this time considering Uranus' oblateness. The global picture does not show a significant difference with the left frame, the main resonant structures remain the same. 

\begin{figure}[h]
\centering
\includegraphics[width=\columnwidth]{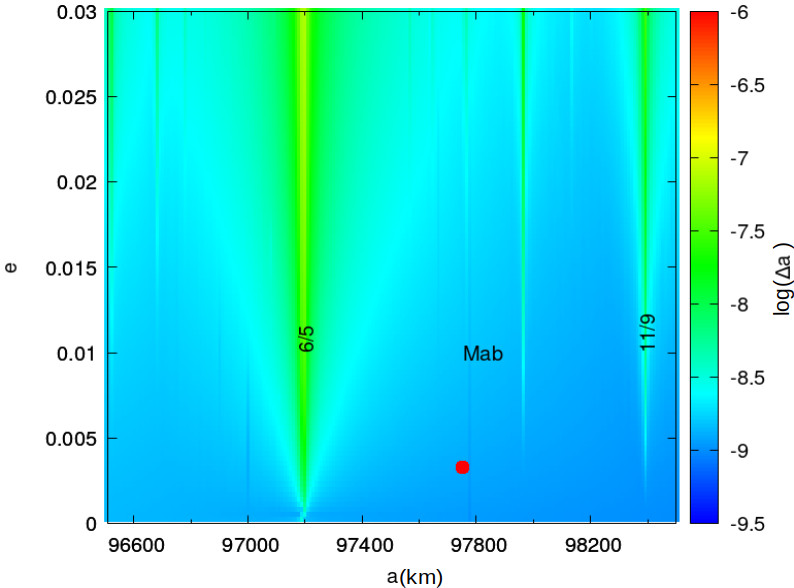}
\caption{Dynamical map in the $(a,e)$ plane around Mab's position using ${\rm max} (\Delta e)$ as the structure indicator. The integration time-span is for $200 \, {\rm yrs}$. 6/5 (left) and 11/9 (right) MMRs with Puck are visible as wider structures at $a=97202 \, {\rm km}$ and $a=98398.6 \, {\rm km}$, respectively. The maps sample 151 x 301 initial $(a,e)$ values around Mab position.}
\label{Bel-Puck-ext}
\end{figure}

We repeat the experiments done in Figure~\ref{Bel-Puck-int} now for the outer region, with $96550 \, {\rm  km} <a< 98350 \, {\rm km}$. Once again, eccentricities were taken between 0 and 0.03, and all angles were set to zero. The integration timespan is 200 years. Results are shown in Figure \ref{Bel-Puck-ext}. The position of Mab ($a \sim 97740 \, {\rm km}$) is denoted with a red circle. The 6/5 and 11/9 MMRs between Mab and Puck are visible on the map. Higher-order resonances can also be seen near the actual position of Mab, although they seem to be very weak for $e<0.01$, Mab's eccentricity (see Table~\ref{tab:a}). 

\subsection{Proximity to 44/43 MMR between Belinda and Perdita}\label{appendix}

\begin{figure}[h]
\centering
\includegraphics[width=\columnwidth]{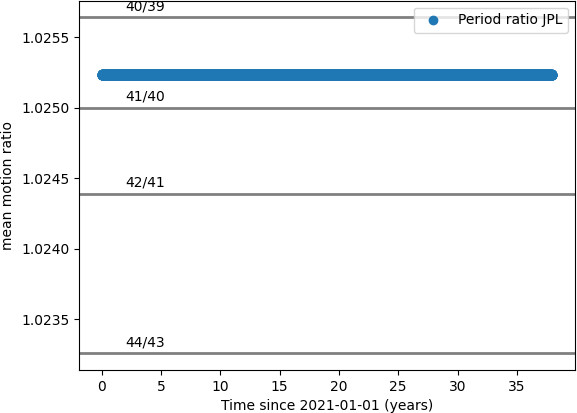} \quad
\includegraphics[width=0.95\columnwidth]{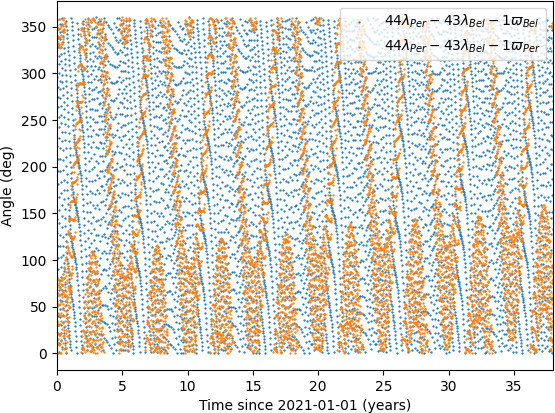}
\caption{\emph{Left panel:} Mean-motion ratio between Belinda and Perdita. \emph{Right panel:} critical angle associated to the not possible 44/43 MMR. \citet[][figure 21, arg2]{French+2015} showed a high amplitude of libration for the angle associated with 44/43 for a time span of $100 \, {\rm yrs}$. Nevertheless, we find that this angle circulates in less than $1\, {\rm yr}$. }
\label{fig:belper}
\end{figure}

\begin{figure}[t]
\centering
\includegraphics[width=\columnwidth]{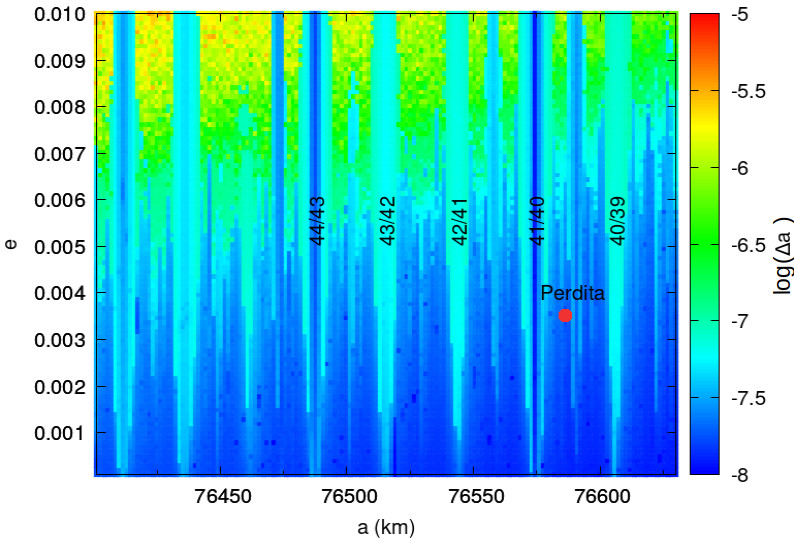}
\caption{${\rm max} (\Delta a)$ dynamical map in ($a,e$) plane considering a system with Belinda, Puck, and third moon in the regions around Perdita location. Initial orbital elements correspond to those retrieved from JPL Horizons and the integration time is $200 \, {\rm yrs}$. Vertical structures depict the resonances in the region (labelled following Figure~\ref{fig:belper}). Some resonances have lower values of $\, {\rm max} (\Delta a)$ at their centre because this is a representative plane with a fixed value of mean anomaly that might not coincide with the libration centres.}
\label{fig:Bel-Puck-intzoom}
\end{figure}

In this section we present a detailed study of the dynamics between Belinda and Perdita. First we retrieve the orbital elements for these moons since 2021/01/01 from JPL Horizons (using the last fit, the URA115 solution) and present the results in Figure~\ref{fig:belper}. The top panel of Figure~\ref{fig:belper}, analyse the period ratio evolution between Belinda and Perdita. The grey lines represent the first-order MMRs near the location of this pair, and in blue we show $n_{\rm Bel}/n_{\rm Per}$. It can be clearly seen that according to the current orbital fits, the pair Belinda/Perdita is not longer at 44/43 MMR. We can see that mean-motion ratio stay almost constant and closer to the 41/40 resonance than to the 44/43. We also calculate the mean-motion ratios from the semi-major axis (i.e, $n_1/n_2 \sim (a_2/a_1)^ {3/2}$) and the results remain unaltered. In the bottom panel of Figure~\ref{fig:belper} we show the ``critical angles'' associated to 44/43 MMR that should be librating if Belinda/Perdita were, in fact, locked in the corresponding MMR, but as can be seen from this plot, it is not the case.
Although Figure~\ref{fig:belper} shows an uneven distribution of the argument $44\lambda_{Per}-43\lambda_{Bel}-\varpi_{Per}$, it might be due to the interaction between the massive moons. The step-size for this integration query is of 2 days, enough time to show that the angle with the new orbital determinations from Horizons does not librate. When comparing with Fig. 21 from \citet{French+2015} and \citet{Quillen+2014} both groups worked with another semi-major axes (older version) that were updated by JPL Horizons. 
This affirmation is sustained by the top panel of Figure~\ref{fig:belper}, where we can see $n_{Bel}/n_{Per}$ far from the nominal location of 44/43 MMR.

Finally, we repeat Figure~\ref{Bel-Puck-int} zooming the region near Perdita's position. We construct a ${\rm max} (\Delta a)$ dynamical map considering Belinda and Puck, while we vary the semi-major axis and eccentricity of a fiducial moon with the mass of Perdita. We set the initial conditions for the orbital parameters and masses retrieved from the JPL and show the results of this integrations in  Figure~\ref{fig:Bel-Puck-intzoom}. As we mentioned before, this region is crowded of resonances, many of them really thin. The current orbital fit for Perdita (identified in the Figure as red dot) is immerse in region surrounded by many weak resonances.
Thus, we show the richness of the resonant structure in the region around the moons and not only rely on a unique best-fit as was done by other authors. The best-fit solution is extremely difficult to obtain, and sometimes errors are usually underestimated, giving for example a covariance matrix. 

Although in \citet{French+2015, French2017} the authors clearly identified the 44/43 MMR between Belinda and Perdita and showed the associated angle librating, with the updated ephemerides and orbital fit published in the JPL, it seems that the system moved apart and that it is no longer inside the resonance.

\section{Speculative paths of dynamical evolution}
Given the proximity to resonant configurations that we identify in the previous section, we analyse the possible past evolution of the outer members of the regular low-mass satellites.

We explore the paths of evolution of satellites within three different scenarios and speculate on some possible histories of the Uranian Satellite System arising either from the interaction with the CPD (gas drag or type-I migration) or tidal evolution between Uranus and the moons.

\subsection{Satellite migration in the CPD}
\label{sect:disc-capture}

Mean-motion resonances play a critical role in sculpting the final structure of a satellite system. We can distinguish this effect by analysing the distribution of the minor bodies of the Solar System, particularly in the sub-systems of satellites around giant planets. Regular satellites display orbits in the same direction of rotation in the equatorial plane as their respective host planets, suggesting the formation process must be analogous to that of the planets around a central star but in a gaseous disc surrounding the planet. 

If satellites formed before the dissipation of the CPD, convergent migration could explain the configuration of the system. Two main mechanisms can generate a convergent migration between satellites: the gas drag and the type-I migration. Both mechanisms usually generate an inward satellite migration. The first one affects small bodies, while the second one massive bodies. The aim of the following sections is to analyse if satellite-disc interactions could lead to a significant satellite migration, but not to compute in detail possible resonant captures between the satellites, due to the fact that we are not considering mutual gravitational interactions when gas drag and type-I migration are considered.

\subsubsection{Migration by gas drag}
\begin{figure}[t]
\centering
    \includegraphics[width=\columnwidth]{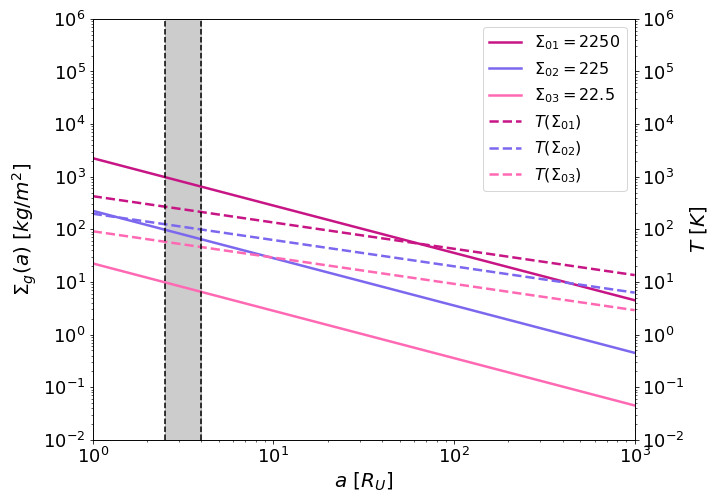}
    \caption{Surface density (solid) and temperature  profile (dashed) for a CPD after $10^4 \, {\rm yrs}$ with our simple approximation for three surface profiles: $\Sigma_{01}$ corresponds to Ida's calculation, $\Sigma_{02}$ and $\Sigma_{03}$ reduced by 10 and 100, respectively. The profile is almost constant for the range of semi-major axis that we study. The shadowed region indicate the location of satellites in study.}
    \label{fig:m-drag-surface-densities}
\end{figure}

In this section, we consider the possible migration or drifting of the satellites from the drag generated by the gaseous component of the CPD. We consider a CPD following \citet{ida2020}, who proposed that the satellite system of Uranus could have been formed in a CPD generated as a consequence of a giant impact. In this sense, these authors showed that after $10^4 \, {\rm yrs}$ of viscous evolution, such CPD reaches a quasi-steady state. At such time, the gas surface density radial profile can be simply approximated by
\begin{equation}
    \Sigma(a) = \Sigma_0 \left(\frac{a}{a_0}\right)^{-\alpha},
    \label{eq1-sec4.1.1}
\end{equation}
with $a_0= \, {\rm R_U}$, $\Sigma_0= 2250 \, {\rm kg/m}^2$ the surface density at $a_0$, and $\alpha= 0.9$. Thus, in Figure~\ref{fig:m-drag-surface-densities}, we show this density profile for the possible disc around Uranus, together with two profiles of $\Sigma$ decreased by a factor 10 and 100. These reductions in the gas surface density tend to mimic a quick dissipation of the CPD in a timescale of $2\times10^4$, as proposed by \citet{ida2020}. Furthermore, in the same figure, we show the temperature profile associated with each $\Sigma_0$. We note that adopting an aspect ratio of $h= 0.1$ \citep{2006Natur.441..834C}, the mid-plane temperature radial profile of the CPD is also similar to the one found by \citet{ida2020} at $10^4 \, {\rm yrs}$. In addition, these authors also showed that at $10^4 \, {\rm yrs}$, the solid material must have condensed beyond $\sim 2 \, {\rm R_U}$ and that the formation of the satellites occurs very quickly by the accretion of the ice condensates (the major satellites can reach masses similar to their current ones in only $\sim 10^3\, {\rm yrs}$). Thus, we study the possible drift of the 6 outer regular satellites, considering their current masses. As the sizes of these satellites are between $\sim 10 \, {\rm km}$ and $\sim 80 \, {\rm km}$ (see Table~\ref{tab:a}), they drift in the quadratic regime, changing their semi-major axis at a rate given by \citet{Adachi1976}
\begin{eqnarray}
    \dfrac{da}{dt}= &\dfrac{2a}{t_{\text{fric}}} \left( \eta^2 + \dfrac{5}{8}e^2 + \dfrac{1}{2}i^2\right)^{\frac{1}{2}}\times \nonumber \\
    &\left\{\eta + \left(\dfrac{5}{16} +  \dfrac{\alpha'}{4}\right)e^2 + \dfrac{1}{4}i^2\right\},
    \label{eq2-sec4.1.1}    
\end{eqnarray}
where $\alpha'$ is the local gradient of the volumetric gas density ($\alpha'= \alpha - 1$ in our model). The eccentricity for each satellite is taken from Table~{\ref{tab:a}}, and we consider for simplicity that the inclinations are half of the eccentricities. The factor $t_{\rm fric}$ is given by
\begin{equation}
    t_\text{fric}= \dfrac{8\rho_{\text{s}}r_{\text{s}}}{3\text{C}_{\text{D}}\rho_{\text{g}}v_{\text{k}}},
\end{equation}
with $\rho_{\text{s}}$ the satellite density --computed as a mean density from the values of the masses and radii given in Table~\ref{tab:a}--, $r_{\text{s}}$ the satellite radius, $\text{C}_{\text{D}}$ a dimensionless coefficient near unity which represents the gaseous friction \citep{Adachi1976}, and $\rho_{\text{g}}$ and $v_{\text{k}}$ the volumetric gas density of the disc and the Keplerian velocity at the location of the satellite, respectively. Finally, the factor $\eta$ is given by
\begin{equation}
    \eta= \dfrac{1}{2} h^2 \dfrac{\text{d} \ln \text{P}_{\text{g}}}{\text{d} \ln a},
\end{equation}
with $\text{P}_{\text{g}}$ the gas pressure of the disc. 

\begin{figure}[t]
\centering
    \includegraphics[width=\columnwidth]{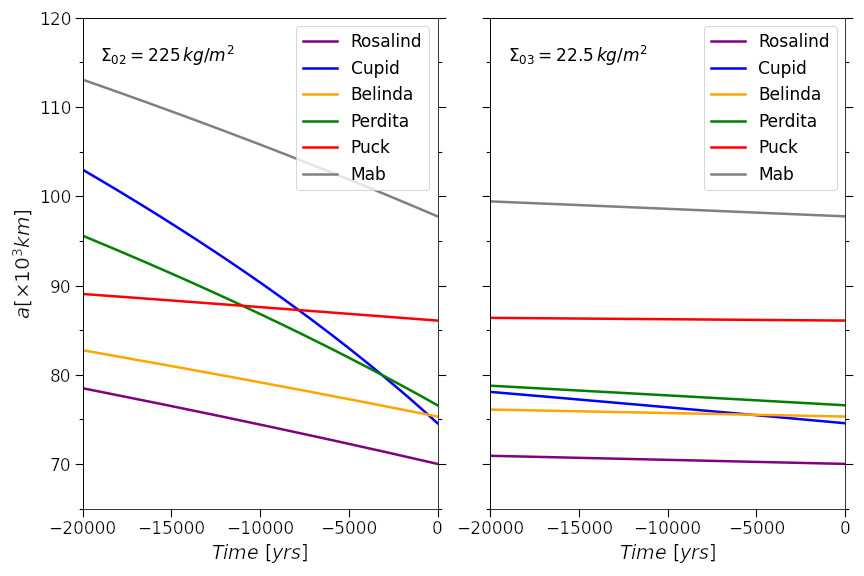}
    \caption{Orbital displacement due to gas drag. The static surface density profiles, $\Sigma_{0i}$ are shown in Figure \ref{fig:m-drag-surface-densities}. Only smaller moons ($R\lesssim 10 \, {\rm km}$) present an appreciable drift in their position considering the disc with profile density $\Sigma_{02}$ (left panel).}
    \label{fig:m-drag}
\end{figure}
\begin{table}[h]
\centering
    \caption{Variation of the semi-major axes due to drag effects on the outer satellites. Values for $\Sigma_{02}=225 {\rm kg/m}^2$ and $\Sigma_{03} = 22.5 {\rm kg/m}^2$ are included. Satellites began the integration in a wider position than currently observed. Thus, drag migration is in the direction to Uranus, indicated in ${\rm kms}$.}
    \label{Tab:drag_mig}
    \begin{tabular}{l|c|c}
    Satellite & $|\delta a| \, (\Sigma_{02})$ & $|\delta a| \, (\Sigma_{03})$ \\
    \hline
    Rosalind & 8689.736 & 936.655 \\
    Cupid    & 29037.736 & 3608.489  \\
    Belinda  & 7616.7137 & 809.688 \\
    Perdita  & 19446.738 & 2254.146  \\
    Puck     & 3063.280 & {313.33} \\
    Mab      & 15698.344 & 1727.616  \\
    \hline
    \end{tabular}
\end{table}

In Figure~\ref{fig:m-drag}, we plot the migration of the satellites due to the gas drag evolution of the CPD, and study two extreme cases, adopting $\Sigma_{02}= 225 \, {\rm kg}/{\rm m}^2$ (left panel), and $\Sigma_{03}= 22.5 \, {\rm kg}/{\rm m}^2$ (right panel). These two cases represent a reduction in one and two orders of magnitude, respectively, with respect to the approximated value of $\Sigma_0$ at $10^4 \, {\rm yrs}$. For the case of $\Sigma_{02}= 225 \, {\rm kg}/{\rm m}^2$, all satellites have a significant migration in just a timescale of $2\times10^4 \, {\rm yrs}$ (except Puck, the major one). For the of $\Sigma_{03}= 22.5 \, {\rm kg}/{\rm m}^2$, just the smaller and closer ones suffer a moderate migration. Thus, we might conclude that if the dissipation of the CPD takes at least 20000 yrs after reaching the quasi-steady-state, gas drag could be a possible mechanism to allow the system to achieve a resonant configuration. In Table~\ref{Tab:drag_mig}, we summarise the variation of the semi-major axis for the surface densities $\Sigma_{02}$ and $\Sigma_{03}$.

It is interesting to note in the left frame of Figure \ref{fig:m-drag} that Rosalinda drifts around 8000 km in the semi-major axis, while Cupid might cross the orbit of Belinda in the last 1000 years of evolution. Also, around 8000 yr in the past, Cupid, Perdita and Puck had very similar semi-major axes. On the other hand, for $\Sigma_0$ decreased by 100, only smaller moons suffered drifts of about $\sim 2000 \, {\rm km}$ ($\sim2\%$) in the last 20000 years of evolution. However, it appears that Cupid had an encounter with Belinda's position $\sim 7000 \, {\rm yrs}$ in the past, perhaps ruling out this scenario for this $\Sigma_0$. 

In both cases there are moons which cross orbits, 4 satellites in the scenario with $\Sigma_{02}$, and 2 in the case with $\Sigma_{03}$. Given the flat disc and size of the moons they would be highly likely to collide at these times, but precise orbital integrations depend on precise parameters of the disc, which are not entirely known. Thus, no scenario can be confirmed or ruled out. However, this computation is useful to see the radial drift of each moon, since we only consider the individual interactions with Uranus, and not the forces between satellites. Let us stress that the treatment is only correct for individual moons because we neglect the mutual gravitational interactions between them. The main goal here was to integrate backwards to recognise changes in the semi-major axis for each moon, although this approach may not give reliable results and individual drift times do not reveal the exact dynamics the moons experienced. However, we find it evident that this effect, although not precise enough, plays an important role at sculpting the final architecture of the system while the CPD is still present.

\subsubsection{Type-I migration}

As satellites grow, gas drag becomes inefficient. However, if satellites become massive enough, they can gravitationally interact with the CPD and generate torques that modifies the orbit of the growing satellite \citep{1997Icar..126..261W,2002ApJ...565.1257T,2004ApJ...602..388T}.

The features of the gaseous disc determine whether the bodies migrate inwards or outwards, at the same time that it dampens or excites the eccentricity and inclination of the planet's orbit. The direction and migration rate depends on the mass of the migrating body and the local physical properties of the gas disc. Although migration is generally inwards, this is not always the case. If the co-rotation torque is dominant \citep{Paardekooper+2011,JimenezMasset2017} or if the thermal torques are included \citep{BL+2015,Masset2017,Guilera2019, Guilera2021} migration can be outward.
 
Satellites form in the CPD, and they will suffer type-I migration because of the satellite-to-planet mass ratio \citep{2002AJ....124.3404C}. Type-I migration affects low-mass bodies which do not open a gap in the disc. The torques exerted on the gas disc by the satellites should not be strong enough to clear their neighbourhood and starve an annular region around their orbit \citep[see][]{Crida+2006,Petrovich2012}.

As happens in the formation of planetary systems, satellites experiment an orbital decay due to the gravitational interactions with the proto-satellite disc. According to \citet{2006Natur.441..834C}, the effects are the same: the presence of the satellite induces spiral waves in the gaseous disc, which inserts a torque on the satellite, making it migrate inward. The authors provide the timescale of this migration for satellites that do not make a gap in the disc, given by 
\begin{equation}
\tau_{a} = \left|\frac{a}{\dot{a}}\right| = \frac{1}{C_a \Omega(a)}
\frac{M_U}{m} \frac{M_U}{\Sigma(a) a^2} \left(\frac{H}{r}\right)^2,
\label{eq:tau-a}
\end{equation}
where $H$ is the vertical thickness of the gas with sound speed $c$. Thus $H/r$ in expression \eqref{eq:tau-a} represents the aspect ratio which it is set fixed to 0.1 \citep[value taken  from][]{2006Natur.441..834C}, $\Omega(a) = \sqrt{G M_U/a^3}$ is the Keplerian angular velocity, and $G$ is the gravitational constant. The expression for $C_a$ is taken from \citet{2002ApJ...565.1257T}, where the analytical formula for the planetary case has been initially proposed for a laminar 3-dimensional isothermal disc, where is given by $C_a=2.7+1.1\alpha$. 

From equation \eqref{eq:tau-a}, if all disc parameters are fixed, the only dependence of the orbital drift duration is with the mass and position of the satellite. Massive moons will have smaller migration timescales, while smaller bodies will have bigger $\tau_a$. This indicates that if two bodies with different masses are in the same radial position, the smaller moon will spiral to the central planet slower than the bigger one.

Inspired by \citet{Peale+1988}, we calculate the possible histories of the exterior inner regular moons  in order to explain their current positions and their passage through other resonances due to satellite-disc interactions. After some algebra, using equation \eqref{eq:tau-a} and following the same prescriptions suggested by \citet{ida2020} given in the previous Section for the surface density profile (see equation \eqref{eq1-sec4.1.1}), we obtain the expression for the variation of the mean-motion,
\be
\frac{d n}{d t} = {\cal C} m n^{-\frac{2}{3}(\alpha-1)}.
\label{dndt_disc}
\ee
Here, ${\cal C}$ depends on all disc parameters, the mass of the migrating satellite, and the mass of the central body, Uranus. 
\begin{equation}
{\cal C} = -\frac{3}{2} C_a \frac{1}{(H/r)^2} \Sigma_0 a_0^\alpha \frac{1}{M_U^2}(GM_U)^{\frac{4}{3}-\frac{\alpha}{3}}    
\end{equation}
gives the complete expression.  
Integrating equation \eqref{dndt_disc}, we obtain
\be
\begin{split}
n &= \left[{\cal C} m \beta (t-t_0) + n_0^\beta  \right]^{1/\beta}, \label{eq:n(t)} \quad \text{where} \\
\beta &= \frac{5-2\alpha}{3} .
\end{split}
\ee
With this last equation, we can calculate the variation of the mean-motion from an initial position $n_0$, since time $t_0$ to time $t$. Note that, as mentioned for the drag effect, these results only considers individual moons, and not gravitational interactions between them.
\begin{figure}[t]
    \centering
    \includegraphics[width=\columnwidth]{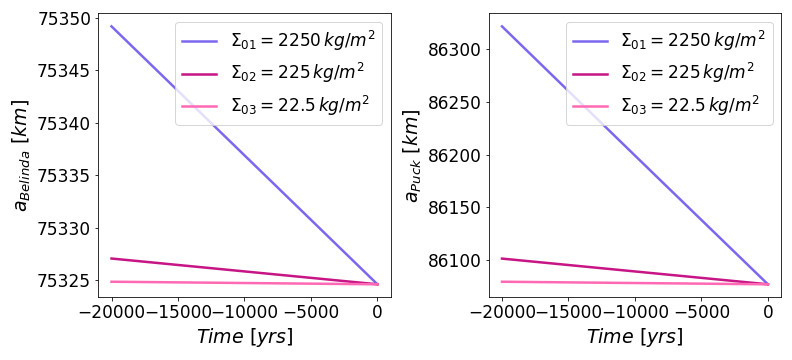}
    \caption{Orbital decay due to type-I migration as a function of time for the two bigger moons of the inner regular low-mass group, Belinda and Puck (see Table~\ref{tab:a}). Each colour represent a different value for the surface density, $\Sigma_{01}=22.5 \, {\rm kg}/{\rm m}^2$ $\Sigma_{02} = 225 \, {\rm kg}/{\rm m}^2$ and $\Sigma_{03}=2250\,{\rm kg}/{\rm m}^2$. }
    \label{fig:a-t_disco}
\end{figure}
\begin{table}[h]
\centering
    \caption{Variation of the semi-major axes of the outer regular moons due to satellite-disc interactions. Extreme values of $\Sigma_0$ are taken for $\alpha=0.9$. Migration is towards Uranus and the displacement is indicated in units of ${\rm km}$.}
    \label{Tab:da_mig}
    \begin{tabular}{l|c|c|c}
    Satellite & $|\delta a| \, (\Sigma_{01})$ & $|\delta a| \,  (\Sigma_{02}) $ & $|\delta a| \,  (\Sigma_{03})$ \\
    \hline
    Rosalind & 12.480 & 1.248 & 0.125 \\
    Cupid    & 0.222 & 0.022 & 0.002 \\
    Belinda  & 24.613 & 2.460 & 0.246 \\
    Perdita  & 0.794 & 0.079 & 0.008 \\
    Puck     & 244.013 & 24.321 & 2.431 \\
    Mab      & 1.214 & 0.121 & 0.012 \\
    \hline
    \end{tabular}
\end{table}

With the results in equation \eqref{eq:n(t)}, we analyse the time it takes for a satellite around Uranus to migrate in discs with different surface density profiles by modifying $\Sigma_0$ while maintaining $\alpha$ fixed in 0.9. As already mentioned, it is known that migration introduced by the disc affects more to higher mass bodies. Therefore, in Figure~\ref{fig:a-t_disco}, we compare the results only for the two bigger moons, Belinda and Puck. 

After converting from mean-motions $n$ to semi-major axis $a$, we show how much Belinda (left) and Puck (right) drifted. $t=0$ shows the position of the satellites in the present, and we integrated backwards for $10^5 \, {\rm yrs}$. In the more favourable situation with $\Sigma_{01}=2250 \, {\rm kg/m^2}$, a small satellite like Cupid moved just 1 km, and a big one like Puck changes its semi-major axis around 1000 kilometres in $10^5 \, {\rm yrs}$. Table~\ref{Tab:da_mig} gives explicit values of the displacement in the semi-major axis for the outer six regular low-mass moons in the last $2\times 10^4 \, {\rm yr}$. Under the best circumstances, Puck is the satellite that exhibits more variation during its formation ($\lesssim 250 \, {\rm km}$), changing at the most $\sim 0.3\,\%$ of its radial position.

Results in Figures~\ref{fig:a-t_disco} and Table~\ref{Tab:da_mig} give complementary information. Migration is extremely low, and hence, the displacement in the semi-major axis is small. We can conclude that Uranus satellites did not experience a considerable orbital decay due to type-I interaction with the circumplanetary disc, meaning that the disc had little effect on the current observed architecture of these small bodies. 

Our results are consistent with those found in \citet{ida2020}, who studied satellite formation for Uranus. However, they studied the classical moons in a more realistic disc. As mentioned above, type-I migration is more important when the bodies are more massive. Therefore, it makes sense that for the smaller moons, the migration effect is also negligible.

\subsection{Tidal evolution of the moons}
\label{sect:tidal}

After the CPD dispersal, tidal interactions between the planet and the small moons became important. The moons of Uranus exhibit some clues about orbital evolution and resonance crossing due to tidal interaction. The major satellites of Uranus revealed surfaces that postdate the final stages of major accretion; those of Ariel and Miranda appear to be especially young \citep{Smith+1986}. The resurfacing of these icy bodies requires a mechanism by which some type of internal energy source is usually needed, and a process of elimination often leads to tidal heating as a last resort \citep{Peale+1988}. A resonance is necessary to maintain significant tidal heating in a synchronously rotating satellite since it forces an orbital eccentricity that would otherwise be rapidly damped. Since low order resonances are not confirmed in the current regular internal satellites, the study of historical resonances could give some clues where tidal heating was a viable mechanism to soften the interiors.

If the satellite orbits had expanded significantly due to torques from tides raised on Uranus, several of the satellite pairs would have reached orbital resonances with a possibility of capture, depending on various values of Uranus' tidal effective dissipation function $Q_U$.  \citet{Peale+1988} considered several possible histories of the satellite system in order to explain the resurfacing events observed in Ariel, while other works studied in detail the passage through other resonances \citep{Tittemore+1988, Tittemore+1989, Tittemore+1990}.  

For two isolated bodies, first-order resonance theory does not provide a mean of disrupting a stable resonance once it is established, so the simplest way to account for the absence of orbital resonances among the satellites today is to assume that the average value of $Q_U$ is sufficiently large ($Q_U > 100000$) that the resonances simply were not encountered over the history of the solar system. However, with this value there is virtually no orbital evolution at all, and no important commensurabilities would have been traversed. 
More important, the minimum value of an average $Q_U$ allows the passage of several important resonances by the Miranda-Ariel, Ariel-Umbriel, and Miranda-Umbriel satellite pairs \citep[see ][and references therein]{Cuk+2020}. \citet{Cuk+2020} determined the tidal dissipation of Uranus using $Q_U = 4 \times 10^4$ (with love number $k_2 = 0.1$), which is roughly the smallest tidal $Q_U$ for which Miranda did not cross the 3:1 MMR with Umbriel, according to \citet{Tittemore+1989}.  

In the following, we describe the tidal model using two values of $Q_U=10000$ and $40000$ and then apply it to the group of satellites we are interested in, to study previous resonances.

\subsubsection{Model}
\begin{table}
\centering
    \caption{Displacement due to tidal evolution assuming two values of $Q_U$, being $Q_1=10000$ and $Q_2=40000$. Negative sign stands for the moons that in the past were closer to Uranus.}
    \label{Tab:Delta-a}
    \begin{tabular}{l|r|r|r}
        Satellite & $a_{\rm 0i}$ & $\delta a_{\rm i1} [{\rm km}]$ & $\delta a_{\rm i2} [{\rm km}] $ \\
        \hline
           Rosalind &	69906.19 & 	 391.3 &    98.9 \\
           Cupid    & 	74372.48 & 	  4.2 &      1.0 \\
           Belinda  &   75235.86 & 	 426.9 &   107.9 \\
           Perdita  &   76397.17 & 	  12.2 &     3.0 \\
           Puck     & 	85985.99 & -1526.4  &  -367.7 \\
           Mab      &	97718.48 & 	  -2.5  &    -0.6 \\
        \hline
    \end{tabular}
    \vspace{-1cm}
\tablecomments{$\delta a_{\rm i1}$ is calculated using $Q_{U1}$ and $\delta a_{\rm i2}$ is calculated \\ using $Q_{U2}$.  }
\end{table}

\begin{figure}[t]
\centering
\includegraphics[width=\columnwidth]{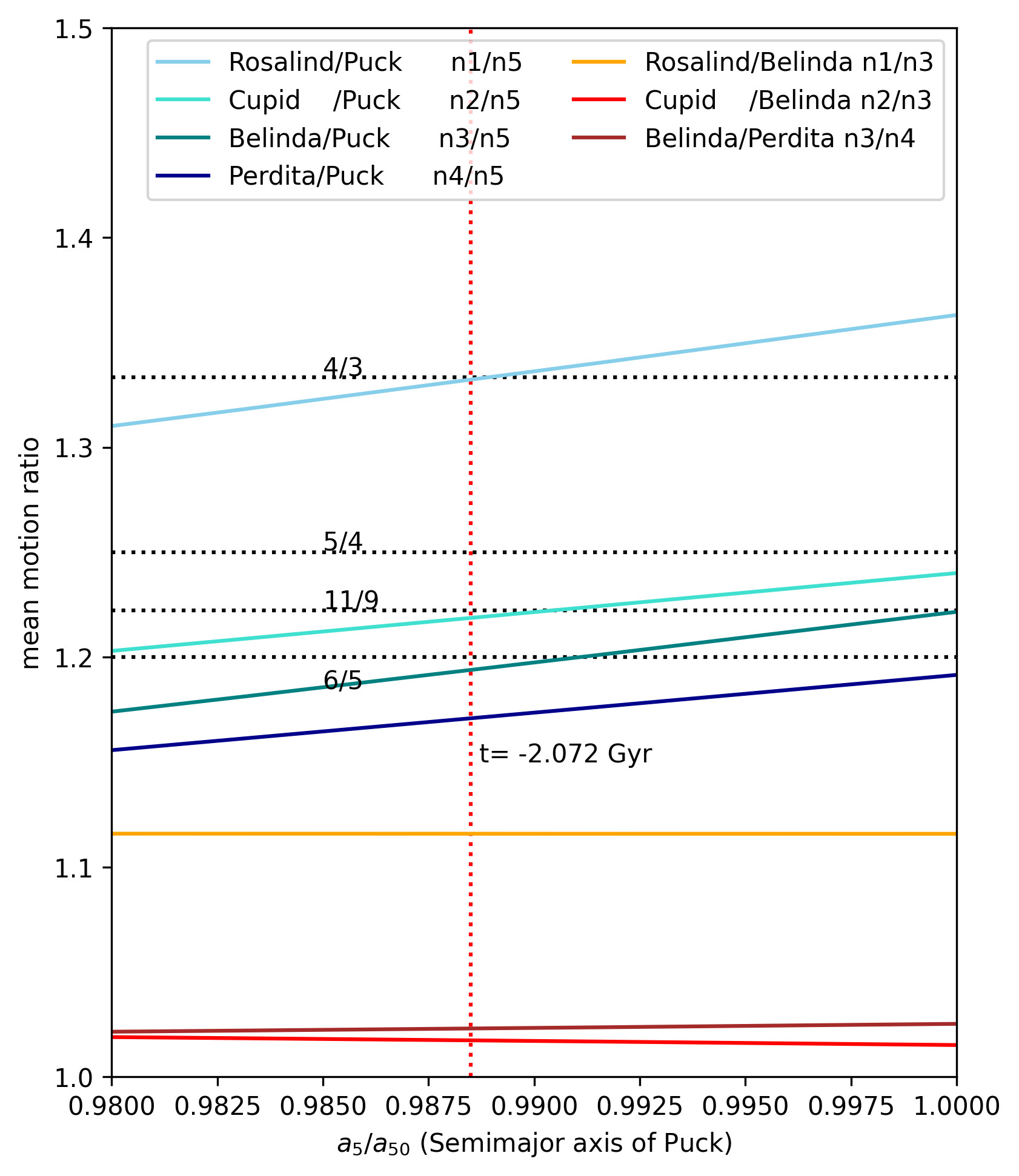}
\caption{Mean-motion ratios of the outer members of the regular internal group, as a function of the change in Puck's semi-major axis ($a_5$), due to tidal dissipation. The subscripts number the satellites from inside out. $a_{50}$ is Puck's current semi-major axis (at $t=0$, where $a_5/a_{50}=1.0$). Puck radially drift 2\% of its orbits, setting an evolution of $4.5 \, {\rm Gyrs}$ with $Q_{U1}=10000$. Horizontal lines shows nominal location of the most important low-order MMRs, also identified in Section \ref{sect:dyn-maps}.}
\label{tidal_evolution}
\end{figure}

We follow the simple model described in \citet{Peale+1988}, assuming that the orbital eccentricities of the Uranian satellites were never very large and that the satellites have the remainder in (or near) the equatorial plane of the planet. The rate of change of the mean orbital angular velocity $n$ from tides raised on Uranus by a satellite that is not in an orbital resonance is given by:
\begin{equation}\label{eq:n}
\frac{dn}{dt}= - \frac{9}{2} k_{2U} \frac{m}{M_U} \frac{R_U^5 n^{16/3}}{Q_U (GM_U)^{5/3}} \text{sgn} (\Omega_U-n)  
\end{equation}
where $k_{2U}$ is the love number for the second-degree spherical harmonic potential, $m$ and $M_U$ are the masses of the satellite and Uranus, respectively. $R_U$ is the radius of Uranus, $a$ is the semi-major axis of the satellite's orbit, $Q_U$ is the specific dissipation function for Uranus, and $G$ is the gravitational constant. The
term $\text{sgn} (\Omega_U-n)$ is equal to 1 if $(\Omega_U-n)$ is positive, and is equal to -1 if it is negative \citep[see e.g.,][]{murray_dermott_1999, Barnes+2002, Sucerquia+2019}. Given the actual rotation period of Uranus ($\sim 0.718$ d), if we consider non interacting moons, all the moons inside synchronous orbit should move inwards because of the tides, and those exterior to the synchronous orbit should move outwards \footnote{For reference, the orbital period of Perdita is $\sim 0.640 \, {\rm days}$, while Puck's is $\sim0.76 \, {\rm days}$. } (those moons beyond Puck).

If we neglect any variation in $Q_U$ or $k_{2U}$ or, equivalently, assign to them average values and assume no stable resonances have existed, equation \eqref{eq:n} can be integrated to yield for the i$-th$ satellite (subscripts numbered from inside out with $i = 1$ for Rosalind, $i = 2$ for Cupid, and so on).
\begin{equation}
n_i^{-13/3} = N m_i (t-t_0) \text{sgn} (\Omega_U-n) + n_{i0}^{-13/3},
\label{eq:mean}
\end{equation}
where
\begin{equation}
N = \frac{13}{3} \frac{9}{2} k_{2U} \frac{1}{M_U} \frac{R_U^5}{Q_U (GM_U)^{5/3}}.     
\end{equation}
The mean-motions with zero subscripts are the current values (alternatively, we can calculate the current semi-major axis). 
$N$ is a positive value and the direction of movement is determined by $\text{sgn} (\Omega_U-n)$. Outside the synchronous orbit, $\text{sgn} (\Omega_U-n)$ takes a positive value, therefore indicating that in the past the moons should have been closer together than in the present (i.e., the orbital drift due to the tidal effect is in the outward direction, and the effect is the opposite of the disc-induced migration). The opposite happens for moons inside Puck's semi-major axis.

\subsubsection{Application to small regular moons}

Assuming a constant value $Q_U$ and $k_{2U}$ we calculate the displacement of the moons we are studying, in the last $4.5 \, {\rm Gyrs}$, using two different values of $Q_U$, and we show the results in Table \ref{Tab:Delta-a}. This displacement represents an upper limit because it considers that the moons where formed in the very beginning of the Solar System.

When considering the first $Q_U$ value fixed in $Q_{U1}=10000$, the variation of Puck's semi-major axis is  $\delta a_5 \sim 1526 \, {\rm km}$, around $2\%$ of its current position, being the moon with major displacement. This quantity reduces to $\sim 0.4\%$ using $Q_{U2}=40000$. Puck and Mab were closer to Uranus 4.5 Gyrs ago ({negative} sign), while the other moons were further than their actual positions. Moreover, Cupid, Perdita and Mab do not experiment an important radial drift by tides. Following \citet{Peale+1988}, a more convenient way to understand past evolution and possible resonance crossing is using the mean-motion ratios ($n_i$/$n_j$) as a function of the change in the semi-major axis of Puck ($a_5$) over the maximum possible range of orbital evolution. 

We can use equation \eqref{eq:mean} to calculate the evolution of the period ratio $n_i/n_j$, considering the pair of moons $i$ and $j$, with $i<j$. Accordingly, we can track the resonance crossing as the pairs evolve. This period ratio allows $k_{2U}/Q_U$ to vary with time as planet properties change, with the only condition that it must have the same value for all the satellites. We choose to show the resonances with respect to Puck and Belinda because they are the ones that have the greater evolution with tides regarding that of the smaller moons. Time decreases as $a_5$/$a_{50}$ decreases (to the left). Our Figure shows the more important resonances identified in the previous Sections, i.e., 5/4, 11/9, and 6/5 MMRs.

Figure \ref{tidal_evolution} show the evolution of the mean-motion ratio as a function of $a_5/a_{50}$. Currently, at $a_5/a_{50}= 1.0$, the pairs Cupid-Puck, Belinda-Puck, and Perdita-Puck are approaching the 5/4, 11/9, and 6/5 MMR, respectively. On the other hand, some pairs of moons tidally evolve, maintaining their mutual distance around a constant value, like the cases of Rosalind-Belinda or Belinda-Perdita (orange and brown lines, respectively). Apparently, back in the past, at $a_5/a_{50} \sim 0.991$, the pair Belinda-Puck crossed the 6/5 commensurability. Furthermore, Cupid-Puck might have also crossed the same resonance for $a_5/a_{50} \sim 0.991$, although unlikely with the assumed actual values of $Q_U=10000$ and $R_U=25600 \, {\rm km}$, this would have happened more than $1 \, {\rm Gyrs}$ ago.

We run many numerical experiments considering the tidal evolution with N-body interactions for the system of Cupid, Belinda, Perdita, and Puck. We use the initial orbital elements of the Uranian satellites from JPL and the equations of motion were solved using an N-body integrator with adaptable step-size and precision of $10^{-13}$ \citep[for details, see ][]{Rodriguez+2013}. We consider mutual tidal interactions in pairs (Uranus and each small moon) using a classical linear tidal model, in which the deformations of the bodies are delayed by a constant tidal time lag $\Delta t$ \citep{Mignard+1979}. We set the initial conditions for the moons in spin-orbit commensurability. However the results change using different precision or even different CPUs (given the chaotic nature of the system). Thus, even with the most recent data for the moons the system is still doomed as was pointed by several authors \citep{French+2012, Quillen+2014}, although some moons are not longer at MMR configurations previously reported.

\section{Discussion and Conclusions}
\label{sec:conclusions}

In this article, we study the dynamics and possible path of evolution of the outer moons within Uranus internal regular satellites, i.e, the small moons outside Rosalind's semi-major axis. We performed dynamical maps the plane $(n_1/n_2,n_2/n_3)$ to understand their location near the two and three-body resonances, and the $(a,e)$ plane to study the multiple structure of resonances.
Additionally, we present diverse dynamical evolutionary scenarios that could have lead these satellites to their current observed configuration, but not analyse the precise ephemerides of the system. We follow pioneering works about Uranus and giant planet formation, like those of \citet{Peale+1988,Showalter+2006,French+2012,Quillen+2014,French+2015,ida2020}.

We find that the moons of Uranus are in the proximity to first and second-order resonances between adjacent and non-adjacent pairs. Many of them are of dynamical interest because they could have played an important role in the past evolution of the system. Belinda and Puck appear to be close to the $11/9$ MMR, Perdita and Puck close to the $6/5$, as well as Puck and Mab. Additionally, it is important to mention that we observe that Cordelia and Ophelia are close to the $9/8$ mean-motion resonance, as well as Ophelia and Bianca close to the $8/7$ commensurability (see Figure~\ref{res-french}). We also included the oblateness effect of the central planet and explored it in different planes, but found no significant difference for the location of the main mean-motion resonances.

The origin of Uranus' proto-satellite disc is still in debate. We follow \citet{ida2020}, who propose that Uranus received a giant impact and its satellite system quickly formed as a consequence of that impact within this disc by the accretion of ice condensates. However, we did not deal with the satellite formation in this work, assuming the small moons were already formed.

We study two main known mechanisms that could move the satellites and might capture them into MMRs, namely disc-driven migration (both by gas drag and classical type-I migration) and tidal effects. 
From the analysis in Section \ref{sect:disc-capture}, we can see that the gas drag effect makes the smaller moons shift radially inwards $\sim 3\times10^4 \, {\rm kms}$ to Cupid and $\sim 2\times10^4 \, {\rm kms}$ to Perdita, in only 20000 years. Type-I migration, on the other hand, is more efficient for the bigger bodies such as Belinda and Puck, that migrate $24$ and $244 \, {\rm kms}$ inwards, respectively, in the same timescale if the disc has superficial density $\Sigma_{01}$ (compatible with results in \citealt{ida2020}).  
After the disc dispersal, tidal interactions start acting. In Section \ref{sect:tidal}, we found that the drift by tidal evolution is not negligible in the past $2 \, {\rm Gyrs}$ for the larger moons: Rosalind, Belinda, and Puck. In fact, this evolution can be tracked to search resonance crossing in the past and maybe in the future. While tidal interactions could have the opposite effect than disc-driven migration for Puck, if the larger moons did not encounter strong MMR, the net effects can compensate. 

Although the interaction with the disc is the classical explanation for the evolutionary process, the in-situ accretion is also a suitable mechanism able to explain near-resonant configurations \citep[see, e.g.,][applied to exoplanet formation]{Petrovich+2013}.
A self-consistent model that takes into account the satellite formation simultaneously with several disc-driven migration mechanisms, as well as mutual interactions, should be conducted to better understand the origin and current configuration of the moons.

\acknowledgments
We thank both Y. Chen and the anonymous reviewer for their reports that helped us improve our manuscript. 

\begin{ethics}
\begin{conflict}
The authors declare that they
have no conflict of interest.
\end{conflict}
\vspace{0.5cm} 

\hspace{-0.5cm}
\textbf{Informed Consent Statement} \\
---

\end{ethics}

\begin{fundinginformation}
CC has been supported by the Fonds de la Recherche Scientifique -- FNRS under Grant No. F.4523.20 (DYNAMITE MIS-project). CC and CG received research grants from CONICET, and Secyt -- Universidad Nacional de Córdoba and used computational resources from CCAD -- UNC which are part of SNCAD (\href{https://ccad.unc.edu.ar}{https://ccad.unc.edu.ar}),  -- MinCyT, República Argentina. OMG is partially supported by the PICT 2018-0934 from ANPCyT, Argentina, and by ANID -- Millennium Science Initiative Program -- NCN19\_171. 
\end{fundinginformation}

\begin{authorcontribution}
The dynamical analysis was made by C. Charalambous and C.A. Giuppone. The analysis about migration was performed by C. Charalambous, the section about gas drag was made by O.M. Guilera and the tidal contribution was made by C.A. Giuppone. \end{authorcontribution}

\begin{dataavailability}
The research done in this project made use of the Astroquery \citep{2019AJ....157...98G}, a community-developed core Python package for Astronomy \citep{2013A&A...558A..33A,2018AJ....156..123A}.
The data presented in this paper is original from the authors, and it is available upon reasonable request.
\end{dataavailability}



\begingroup
\let\clearpage\relax
\bibliographystyle{apsrev}
\bibliography{biblio} 

\endgroup

\end{document}